# Ferromagnetism in laser deposited anatase Ti$_{1-x}$Co$_x$O$_{2-\delta}$ films.

S. R. Shinde[*] and S. B. Ogale[+]

Center for Superconductivity Research, Department of Physics, University of Maryland, College Park, MD 20742-4111.

S. Das Sarma[§], J. R. Simpson and H. D. Drew

Department of Physics, University of Maryland, College Park, MD 20742-4111.

S. E. Lofland and C. Lanci

Department of Chemistry and Physics, Rowan University, Glassboro, N.J. 08028-1701

J. P. Buban and N. D. Browning

Department of Physics, University of Illinois at Chicago, 845 West Taylor Street, Chicago, IL 60607-7059.

V. N. Kulkarni[¶], J. Higgins, R. P. Sharma, R. L. Greene and T. Venkatesan

Center for Superconductivity Research, Department of Physics, University of Maryland, College Park, MD 20742-4111.

[*] shinde@squid.umd.edu
[+] Also at the Department of Materials Science and Nuclear Engineering, ogale@squid.umd.edu.
[§] Condensed Matter Theory Center
[¶] On leave from Indian Institute of Technology, Kanpur, India.

**Abstract**

Pulsed laser deposited films of Co doped anatase $TiO_2$ are examined for Co substitutionality, ferromagnetism, transport, magnetotransport and optical properties. Our results show limited solubility (up to ~ 2 %) of Co in the as-grown films and formation of Co clusters thereafter. For $Ti_{0.93}Co_{0.07}O_{2-\delta}$ sample, which exhibits a Curie temperature ($T_C$) over 1180 K, we find the presence of 20-50 nm Co clusters as well as a small concentration of Co incorporated into the remaining matrix. After being subjected to the high temperature anneal during the first magnetization measurement, the very same sample shows a $T_C$ ~ 650 K and almost full matrix incorporation of Co. This $T_C$ is close to that of as-grown $Ti_{0.99}Co_{0.01}O_{2-\delta}$ sample (~700 K). The transport, magnetotransport and optical studies also reveal interesting effects of the matrix incorporation of Co. These results are indicative of an intrinsic $Ti_{1-x}Co_xO_{2-\delta}$ diluted magnetic semiconductor with $T_C$ of about 650 – 700 K.

**Introduction**

Diluted magnetic semiconductors (DMS) have attracted considerable recent attention[1] because of their potential for novel applications (combining magnetics, electronics, and photonics) in the rapidly evolving area of spintronics. Equally important is the fundamental issue of the origin and the nature of ferromagnetism in these low carrier density semiconductor systems. While carrier induced interaction between the magnetic atoms (e.g. Mn, Co) is suggested as the important ingredient underlying ferromagnetism in the DMS, the precise mechanism is still controversial and is being actively debated in the literature[2].

Recent work in this area has focused mostly on Mn doped III-V compound semiconductors, e.g. $Ga_{1-x}Mn_xAs$. Subsequently, however, the search has been extended to materials such as the Mn doped group IV semiconductors (e.g. $Mn_xGe_{1-x}$) and various oxide systems in view of the rich property-space the oxides have to offer. In the oxide family, DMS ferromagnetism was first reported in Co doped ZnO via magnetoresistance and magnetic circular dichroism measurements[3,4]. Ueda et al.[5] reported a Curie temperature ($T_C$) as high as 300 K in some of their $Zn_{1-x}Co_xO$ films, although the reproducibility was less than 10%, with most samples exhibiting spin glass behavior, raising questions about the intrinsic nature of ferromagnetism in $Zn_{1-x}Co_xO$. Recently, Matsumoto et al.[6] reported occurrence of ferromagnetism above room temperature in Co doped anatase $TiO_2$, which generated great interest in the condensed matter community, creating new multifunctional possibilities for materials design and exploration using oxide based ferromagnetic semiconductor materials. Some significant advantages of ferromagnetic Co:$TiO_2$ are: a) its reported high temperature ferromagnetism which makes it particularly attractive from a technological perspective, b) the relative ease of materials synthesis and processing, and c) the electronic nature of the carrier system (in the Mn-doped III-V DMS systems the carriers are holes making it problematic for actual applications).

Anatase $TiO_2$ is a wide band gap semiconductor with excellent optical transmission in the visible and far infrared regions, a high refractive index, a high dielectric constant, and useful photocatalysis properties. It has been shown to have a very shallow donor level, high n-type mobility, and a large thermopower. The n-type carriers (i.e. electrons) in $TiO_2$ result from oxygen vacancies, which provide tunability to its properties. The few studies[6,7] performed on anatase $Ti_{1-x}Co_xO_2$ thus far suggest the following: (i) Co is soluble in anatase $TiO_2$ at least up to x = 0.08 – 0.1, with no obvious indication of clustering; (ii) $T_C$ is well above 400 K; (iii) the oxidation state of Co is 2+; and (iv) the films exhibit a large magnetoresistance at very low temperatures and high fields. Unfortunately however, there is neither a clear determination of $T_C$ for the system, nor a specific model which could bear out a significantly high $T_C$. This has lead to suggestions/speculations about Co clustering and the very interest in this and similar systems hinges on the resolution of these matters.

In this paper, we examine pulsed laser deposited high quality films of Co doped anatase $TiO_2$ for their various physical properties. For the $Ti_{0.93}Co_{0.07}O_{2-\delta}$ sample grown by pulsed laser deposition (PLD), which shows a high $T_C$ of over 1180 K, we do find evidence of several 20-50 nm Co clusters, with a small concentration of dilutely dispersed Co in the matrix. However, upon high temperature treatment we find almost full incorporation of Co and a $T_C$ ~ 650 K. In fact, the low doped ($Ti_{0.99}Co_{0.01}O_{2-\delta}$) system shows a $T_C$ ~ 700 K in the as-grown state itself. These results indicate that Co located on ordered lattice sites in $TiO_2$ leads to a $T_C$ of 650 – 700 K, showing this to be one of the highest $T_C$ DMS systems.

**Experimental**

Thin films of $Ti_{1-x}Co_xO_{2-\delta}$ with x = 0, 0.01, 0.02, 0.04, 0.07, and 0.15, were deposited on $SrTiO_3$ (001) and $LaAlO_3$ (001) substrates by PLD, using sintered targets synthesized by the standard solid-state route. The substrate temperature, laser energy density, and pulse repetition rate were kept at 700 °C, 1.8 J/cm², and 10 Hz, respectively. Significant room temperature magnetization was found for films deposited at an oxygen pressure in the range between 1 x $10^{-5}$ to 1 x $10^{-4}$ Torr. Most results reported here are for 2000 Å thick films deposited at a pressure of 2-4 x $10^{-5}$ Torr. However, for $T_C$ measurement using vibrating sample magnetometry (VSM), thicker films (1 - 6 μm) had to be used in order to obtain good signal to noise ratio. We carefully ensured that the thick samples had the same magnetic, structural, and electrical properties as the thin films used in the other experiments. X-ray diffraction (XRD) and Ion Channeling techniques (with 1.5 MeV $He^+$ ions) were employed for structural characterization. The hysteresis loops and magnetization were obtained using a SQUID magnetometer at temperatures from 5 to 350 K. The standard four-terminal geometry was used to obtain the electrical resistivity ($\rho$). The microscopic analysis was performed using a JEOL 2010F 200kV field emission scanning transmission electron microscope (STEM). This microscope is capable of generating a 0.14 nm probe size for Z-contrast imaging and electron energy loss spectroscopy (EELS)[8]. The advantage of the STEM analysis over conventional transmission electron microscopy (TEM) is that it provides a direct image of compositional variations and the spectroscopy allows valence states to be identified.

**Results and Discussion**

Fig. 1(a) shows a representative $\theta - 2\theta$ XRD pattern corresponding to a $Ti_{0.93}Co_{0.07}O_{2-\delta}$ film on $SrTiO_3$ (001). The Co concentration (7 at.%) used in this case is the same as that used by Matsumoto et al.[6] and Chambers et al.[7]. The film peaks can be clearly identified with the (00l) planes of the anatase phase. The width of the XRD rocking curve (shown in the inset) for these films is ~ 0.3°, which is comparable with that of the single crystal substrate (0.2°) as well as a pure $TiO_2$ film (0.3°). However, the intensity of the film peaks is substantially lower than that for a pure anatase $TiO_2$ film deposited under the same condition, which suggests presence of matrix disorder. In Fig 1(b) we show the hysteresis loops at 300 and 5 K for this case. In the inset of Fig. 1(b) is shown the 300 K hysteresis loop on the expanded scale. This confirms the room temperature ferromagnetism in the system as observed by Matsumoto et al.[6].

The Ion Backscattering Channeling data for pure anatase $TiO_2$ and the $Ti_{0.93}Co_{0.07}O_{2-\delta}$ films are summarized in Fig 2. The observed minimum yield, $\chi_{min}$, of ~ 5% for channeling in pure $TiO_2$ reflects very high crystalline quality of the undoped film [Fig. 2(a)], thereby indicating optimum growth conditions for $TiO_2$. The channeling for Ti in the $Ti_{0.93}Co_{0.07}O_{2-\delta}$ film [Fig. 2(b)], however, is relatively poor ($\chi_{min}$ ~ 25%). Also, the channeling in the Co contribution to the spectrum [Fig. 2(c)] is almost (i.e. within the accuracy of about 20% of the analysis procedure) absent. This was further confirmed by performing the RBS channeling study at higher ion energy of 3.05 MeV, which clearly separates out the Co signal; the result being shown in Fig. 2(d). These observations suggests three possibilities: Co interstitials (either isolated or in the form of agglomerates), structurally incoherent clusters of Co-Ti-O complex, or Co metal nanoclusters[9]. We note that these possibilities inflict a negative distortive influence on the surrounding $TiO_2$, which is consistent with the reduced x-ray diffraction peak intensity.

In Fig. 3 we show the results of magnetization measurements for the $Ti_{0.93}Co_{0.07}O_{2-\delta}$ films. The SQUID data obtained for 5-350 K shows that the saturation magnetization is almost constant over this temperature range, in complete agreement with the previous reports[6,7]. The saturation magnetic moment in our films is however found to be ~ 1.4 $\mu_B$/Co, which is higher as compared with that of Matsumoto et al.[6]. We note that our oxygen pressure during deposition was higher than that employed by Matsumoto et al.[6]. Chambers et al.[7] used oxygen-plasma-assisted molecular beam epitaxy to grow $Co_xTi_{1-x}O_2$ films at low growth rate and oxygen-rich conditions, which could possibly lead to a different state of the sample. Interestingly however, their magnetization (~1.26 $\mu_B$/Co) is comparable to ours. A nearly constant magnetization over 5-350 K range was also observed for other dopant concentrations between 1-15% Co, indicating that the $T_C$ for all such films is well above 350 K.

We employed VSM to obtain the $T_C$. Figure 3 also shows these magnetization data measured in pure argon ambient up to 1200 K. A clear transition is seen near 1180 K, which is remarkably high and close to the $T_C$ of pure Co metal. This strongly suggests the presence of Co metal clusters in the film, which we establish by STEM and EELS, as discussed below. We also like to point out that the hysteresis loops measured at low temperature under zero field cooled (ZFC) and field cooled (FC) conditions (shown in the inset) did not exhibit any significant relative shift, implying the absence of any aniferromagnetic skin layer such as CoO and the attendant exchange coupling effects. We further note that the magnetization as a function of temperature as measured by VSM is not characteristic of a typical ferromagnet and shows some structure, especially around 700-800 K.

In Fig. 4 we summarize the key STEM results from the $Ti_{0.93}Co_{0.07}O_{2-}$ sample. Several clusters having size of 20-50 nm can be immediately seen in the cross-sectional image shown in Fig. 4(a). EELS data [Fig. 4(b)] acquired from each cluster confirm a large concentration of Co in these clusters. There are some Ti and O signals in four of the clusters, but we believe this to be due to an overlaying of the Co cluster on the $TiO_2$ bulk. This is confirmed by the EELS spectrum from the cluster indexed #5, which is on the edge of the specimen and not on top of any $TiO_2$ layers. This cluster shows a Co signal with no oxygen or titanium peaks. There is a weak Co signal away from the clusters, indicating the presence of some Co in the bulk of the $TiO_2$ matrix. From the STEM images we cannot see any signatures of interstitials (i.e. no large strain fields in the images), suggesting the possibility of a small quantity of Co occupying either substitutional sites or the natural Ti vacancy sites in anatase (anatase has alternate Ti removed from $Ti_2O_2$ matrix). In other words, the images suggest that some Co resides in the actual lattice columns and not between. We also obtained high-resolution plan-view images [Fig. 4(c)] of the regions away from the clusters. The Z-contrast images are rather uniform and no strain fields are apparent which could suggest interstitials. Some bright spots are encountered which could suggest the incorporation of a high Z element such as Co, although further analysis is required to verify that these bright spots are not caused by surface inhomogeneities. All the STEM data thus show the presence of Co clustering in the film as well as incorporation of a small fraction of Co in the remaining $TiO_2$ matrix.

It is now useful to consolidate the emerging picture of a portion of Co being incorporated and the remaining being clustered in the as-grown films by the study of lattice relaxation effects as a function of Co concentration. In Fig. 5 we show the evolution of the (00l) lattice parameter as a function of Co concentration. It is clearly seen that the lattice parameter shows a concentration dependent relaxation at low concentrations up to about 1.5 - 2 % and a saturation behavior at higher concentrations, suggesting clustering of excess Co. The intensity of the main (001) reflection of the anatase phase as a function of Co concentration (inset) also implies that Co incorporates well into the matrix at low (< ~2%) concentration, but forms clusters at higher concentrations, under our growth conditions.

In order to explore whether the high temperature VSM measurement causes any changes in the state of the sample, we remeasured the same sample of Fig. 3 up to high temperatures. The corresponding data are shown in Fig. 6 (a). Remarkably, a clear magnetic transition is observed with a $T_C$ near ~ 650 K, and there is hardly any magnetic moment at higher temperature. The room temperature moment in the remeasured (and therefore annealed) sample is found to be close to that in the as-grown sample within experimental accuracy. This shows that most of the Co is still present and is in the ferromagnetic state even after the high temperature treatment. No additional peaks are observed in the XRD spectrum of the high temperature treated film and the $TiO_2$ peak intensity is seen to increase after the heat treatment. These observations suggest uniformly incorporated Co in an anneal-induced structurally improved matrix with a $T_C$ of ~ 650 K. Interestingly this is near the temperature of ~ 700 K at which a peculiar structure is seen in the magnetization of the as-grown film. This implies that the high temperature treatment has lead to an almost complete matrix incorporation of Co. Our ion backscattering channeling data [inset of Fig. 6(a)] confirm this, with the minimum yield of Co in the annealed case found to be ~ 30 ± 5 %, as against almost no channeling for the as-deposited case. The STEM image shown in Fig. 6(b) also corroborates this with no indication of Co metal clusters. The EELS analysis however indicated that the concentration of the incorporated Co was not entirely uniform across the film. Further work is being done to explore whether this is due to incomplete annealing or is an intrinsic feature. Finally, It is interesting to point out that the lattice parameter value for the annealed samples (Fig. 5, shown as filled circle and open square) falls on the same dashed curve, which represents the concentration dependence of lattice parameter below the saturation regime. This further supports that after high temperature treatment even 7% Co gets matrix incorporated with the corresponding lattice relaxation.

The suggestion of a potentially intrinsic DMS with $T_C \sim 650$ K encouraged us to also examine the high temperature magnetization of a low doped material, namely $Ti_{0.99}Co_{0.01}O_{2-}$ , wherein, *a priori*, clustering effects can be expected to be far weaker even in the as-grown state. We recall from Fig. 5 that at such low Co concentrations clear lattice relaxation was observed implying Co incorporation into the matrix without Co metal cluster formation. The corresponding magnetization data [Fig. 7 (a)] indeed establish the existence of a transition near 700 K, even for the as-grown film. The XRD data [inset of Fig. 5] for this low doped sample shows an intensity almost comparable to that of undoped anatase $TiO_2$ film. The STEM results [Fig. 7(b) and (c)] show no indication of the presence of any Co metal clusters.

In Fig. 8 we compare the transport and magnetotransport data for the undoped and $Ti_{0.99}Co_{0.01}O_{2-}$ films. Comparison of the resistivity dependence on temperature [Fig. 8(a)] for these two films brings out that the influence of incorporated Co on the transport in anatase is quite significant. The Hall effect measurements show n type conduction in both films with estimated carrier concentrations of $\sim 1.4 \times 10^{18}$ /$cm^3$ and $\sim 2.1 \times 10^{18}$ /$cm^3$, respectively at 300 K. Conduction in anatase $TiO_2$ has been suggested to be due to the presence of oxygen vacancies whereby states are introduced just below the conduction band[10]. These states are depopulated at only a few Kelvin as reflected by an upturn in the resistivity at low temperature for the undoped film. The conduction at temperature above a few Kelvin is basically band-like. The change in this nature of resistivity of pure anatse $TiO_2$ caused by Co incorporation can occur by its connection to vacancy population via its need to satisfy its valence ($Co^{2+}$) as well as its contribution to magnetic scattering. Increase in resistivity in spite of an increase in carrier concentration implies drop in mobility due to magnetic or strain scattering effects. The effect of magnetic contribution of Co to transport is evidenced more clearly in the low temperature magnetoresistance (*MR*, defined as $[R(H)-R(0)]/R(0)$, where $R(H)$ and $R(0)$ are the resistance values with and without applied magnetic field, respectively) data shown in Fig. 8 (b). The *MR* is positive and is significant only at very low temperatures which correspond to conditions representing partial ionization of shallow donor states, generally attributed to oxygen vacancies. It shows an approximately quadratic dependence on applied field. While the *MR* in undoped film at 3K in a field of 8 T is about 6 %, it is as high as 23 % for the $Ti_{0.99}Co_{0.01}O_{2-}$ film and 40% for the $Ti_{0.98}Co_{0.02}O_{2-}$ under comparable conditions. For higher Co concentrations the *MR* also shows saturation due to clustering of excess Co in the as-grown state, as discussed earlier. Based on the understanding that the oxygen vacancy related states lie close to the bottom of the conduction band, the observed large low temperature positive *MR* in the $Ti_{0.99}Co_{0.01}O_{2-}$ film can be attributed to Zeeman splitting of this band of states through their coupling to Co spin. Since the lower split band will be occupied the *MR* should be positive, as seen. This feature also highlights the significance of the combined role of the magnetic atom and a defect state (vacancy) in controlling the physical and possibly the magnetic properties.

Finally, in Fig. 9 we compare the optical data on the undoped and 2% Co doped anatase $TiO_2$ films. The absorption edge shows a clear blue shift signifying the influence of incorporated Co on the electronic states. Such shifts have been found in diluted magnetic semiconductor systems[11,12]. More detailed work on the optical conductivity as a function of Co concentration with and without annealing is now in progress and will be reported separately.

**Conclusion**

Pulsed laser deposited films of Co doped anatase $TiO_2$ are shown to exhibit ferromagnetism at high temperature in the as-grown state, with coexisting contributions of Co metal clusters, and that of dispersed matrix-incorporated Co. High temperature treatment is shown to enhance the incorporation of Co into the matrix dramatically. The as grown low-doped (~ 1% Co) case, as well as the annealed high-doped (7%) case, are indicated to be intrinsic diluted magnetic semiconductor (DMS) systems with a high Curie temperature of ~ 650 - 700 K. Transport, magnetotransport and optical properties confirm the matrix incorporation of Co at low concentrations even in the as-grown state. Changes in the matrix incorporation of Co upon high temperature treatment without significant loss of the magnetic moment is encouraging from the standpoint of search for new growth conditions for enhanced incorporation of Co in as-grown state.

**Acknowledgement**

This work was supported by NSF under the MRSEC grant DMR-00-80008, by DARPA (T.V., S.B.O.) by US-ONR (S.D.S.) and DARPA (S.D.S.), and New Jersey Commission on Higher Education (S.L.). Two of the authors (J.P.B. and N.D.B.) would like to acknowledge DOE funding under grant DE-FG02-96ER45610 and NSF support for the purchase of the JEOL microscope under grant DMR-9601792. We acknowledge fruitful discussions with Sang-Wook Cheong (Rutgers), H. D. Drew (UMD), J. Gopalakrishnan (IISc, Bangalore), and A. J. Millis (Columbia). We also thank Robin Farrow (IBM) and Scott Chambers (PNL, Washington) for sharing their unpublished data with us and their comments.

**Figure Captions:**

Fig. 1: (a) The $\theta - 2\theta$ XRD spectrum for a $Ti_{0.93}Co_{0.07}O_{2-\delta}$ film on $SrTiO_3$ (001). Peaks labelled "S" correspond to substrate. The inset shows the XRD rocking curve of the same film. (b) Magnetization (*M*) as a function of temperature (*T*) for a $Ti_{0.93}Co_{0.07}O_{2-\delta}$ film. The 300 K hysteresis loop is shown on the expanded scale in the inset.

Fig. 2: (a) The 1.5 MeV $He^+$ Rutherford Backscattering (RBS) random and channeled spectra for a pure $TiO_2$ film on $LaAlO_3$ (001). (b) The 1.5 MeV $He^+$ RBS random and channeled spectra for a $Ti_{0.93}Co_{0.07}O_{2-\delta}$ film. (c) The Ti and Co contributions obtained by subtraction of the substrate La contribution from the spectra of Fig. 2(b). (d) The 3.05 MeV $He^+$ RBS random and channeled spectra for the film of Fig. 2(b). The use of higher energy clearly separates out the Ti and Co contributions. The sharp peak seen near channel 160 corresponds to the resonant scattering from oxygen.

Fig. 3: The *M-T* data for a $Ti_{0.93}Co_{0.07}O_{2-\delta}$ film. The inset shows the hysteresis loops obtained under ZFC and FC conditions.

Fig. 4: (a) Cross-section STEM dark field image of a $Ti_{0.93}Co_{0.07}O_{2-\delta}$ film. (b) EELS data recorded for the cluster regions marked in STEM image. (c) High resolution plan view image of a region away from the clusters.

Fig. 5: The (004) lattice spacing as a function of Co concentration for films deposited at two different pressures. The filled circle and open square correspond to the annealed $Ti_{0.93}Co_{0.07}O_{2-\delta}$ films. The dashed curve is a guide to the eye. The inset shows normalized intensity of the anatase (004) XRD peak as a function of Co concentration.

Fig. 6: (a) The *M-T* data for the film of Fig. 3 remeasured after high temperature treatment during the first measurement. The RBS random (line) and channeled (circles) spectra after subtraction of the substrate La contribution for the annealed $Ti_{0.93}Co_{0.07}O_{2-\delta}$ film are shown in the inset. For comparison the channeled spectrum (squares) for the as-grown film is also shown. (b) Cross-section STEM dark field image of the $Ti_{0.93}Co_{0.07}O_{2-\delta}$ film after annealing.

Fig. 7: (a) The *M-T* data for the $Ti_{0.99}Co_{0.01}O_{2-\delta}$ film. (b and c) Cross-section STEM dark field images of $Ti_{0.99}Co_{0.01}O_{2-\delta}$ as-grown film.

Fig. 8: (a) Resistivity as a function of temperature for $TiO_2$ and $Ti_{0.99}Co_{0.01}O_{2-\delta}$ films. (b) The magnetoresistance as a function of magnetic field for $TiO_2$ and $Ti_{1-x}Co_xO_{2-\delta}$ films.

Fig. 9: Optical transmission spectra for the $Ti_{1-x}Co_xO_{2-\delta}$ films with x= 0, and 0.02.

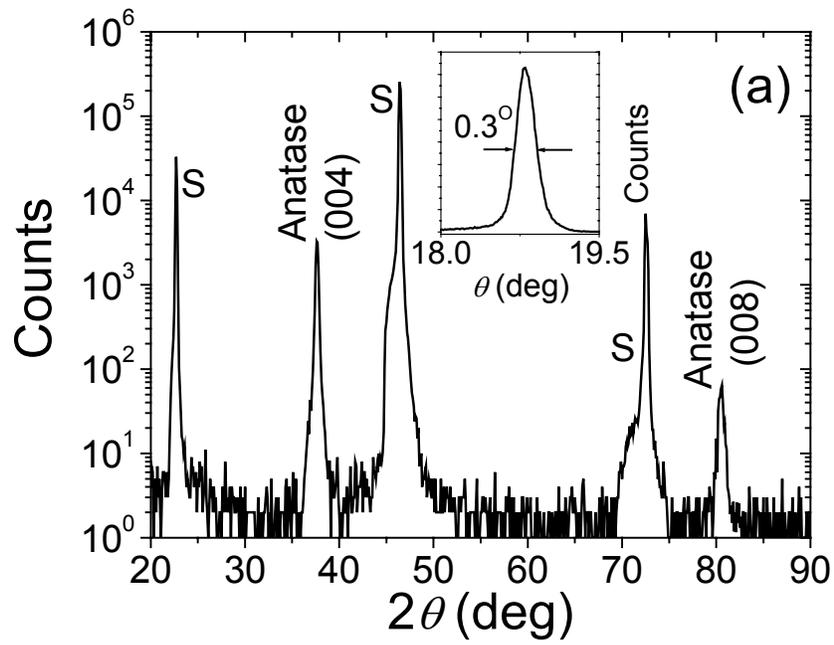

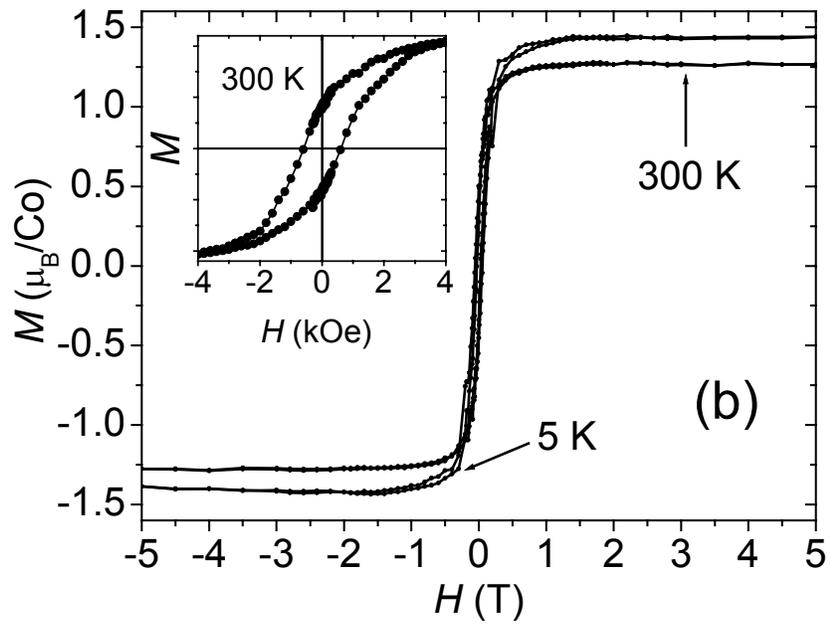

Fig. 1: S.R. Shinde *et al*.

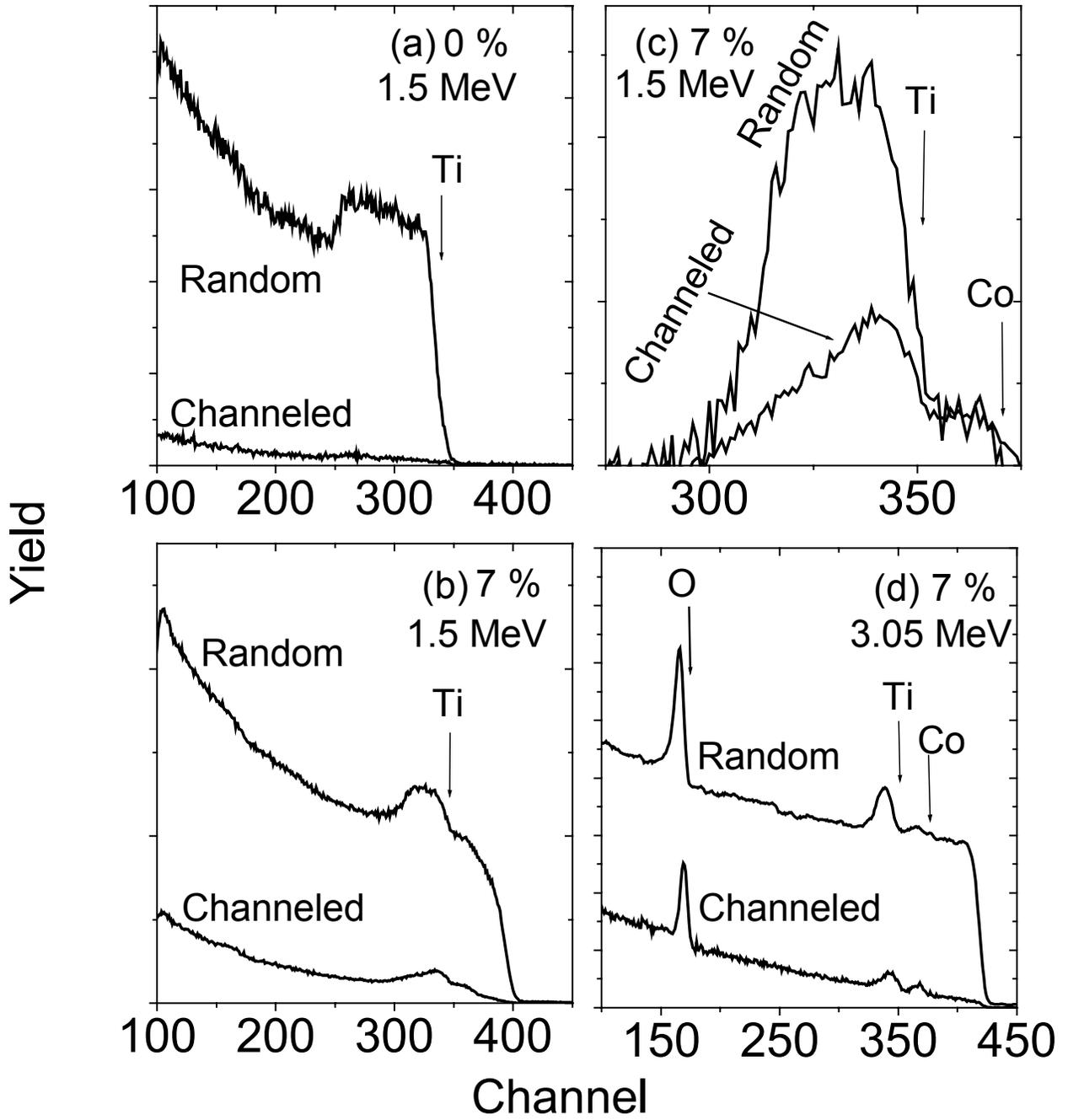

Fig. 2: S.R. Shinde *et al*.

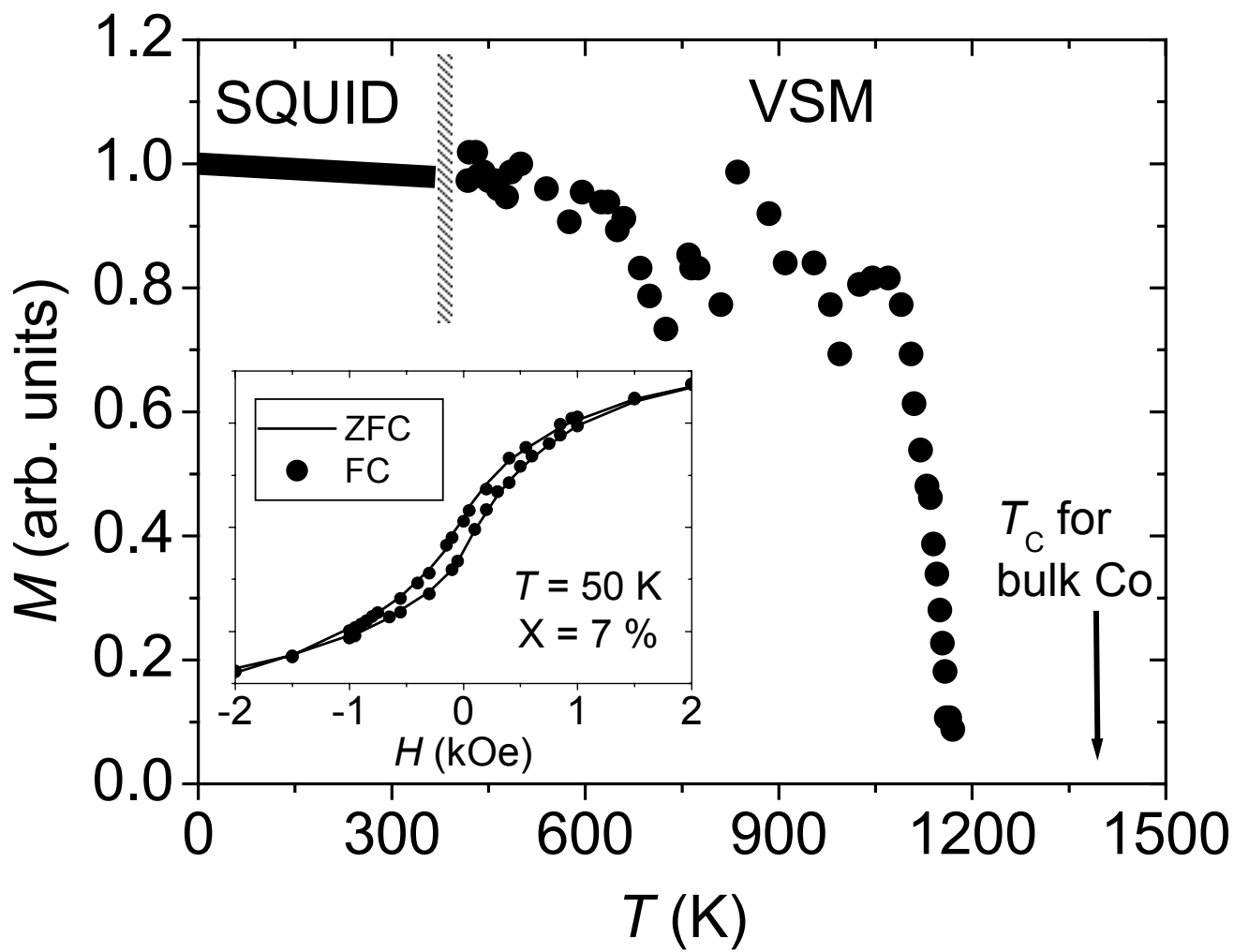

Fig. 3: S.R. Shinde *et al*.

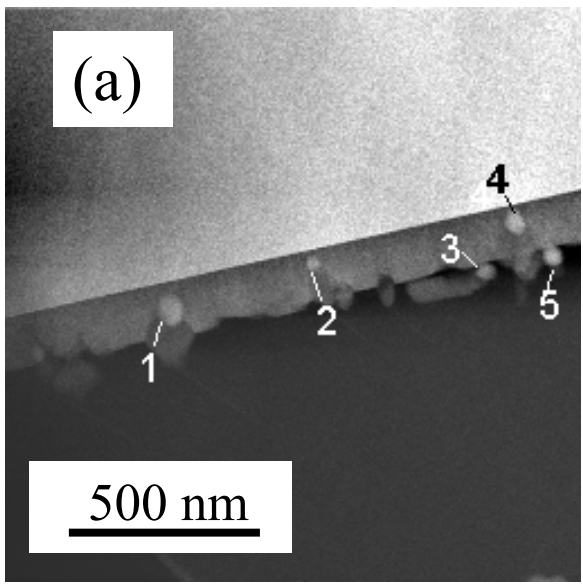
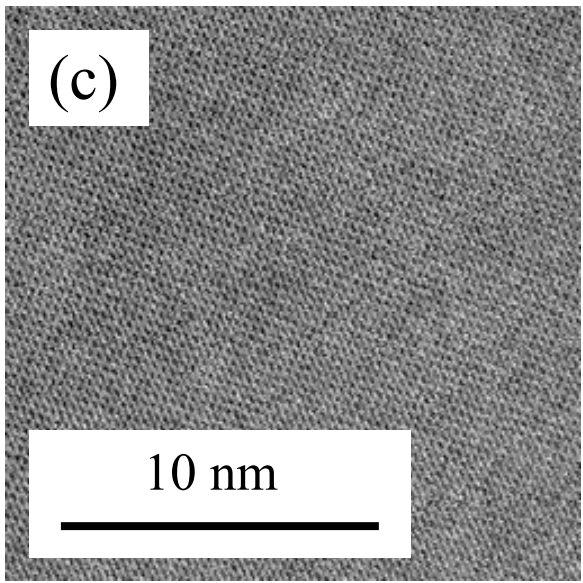
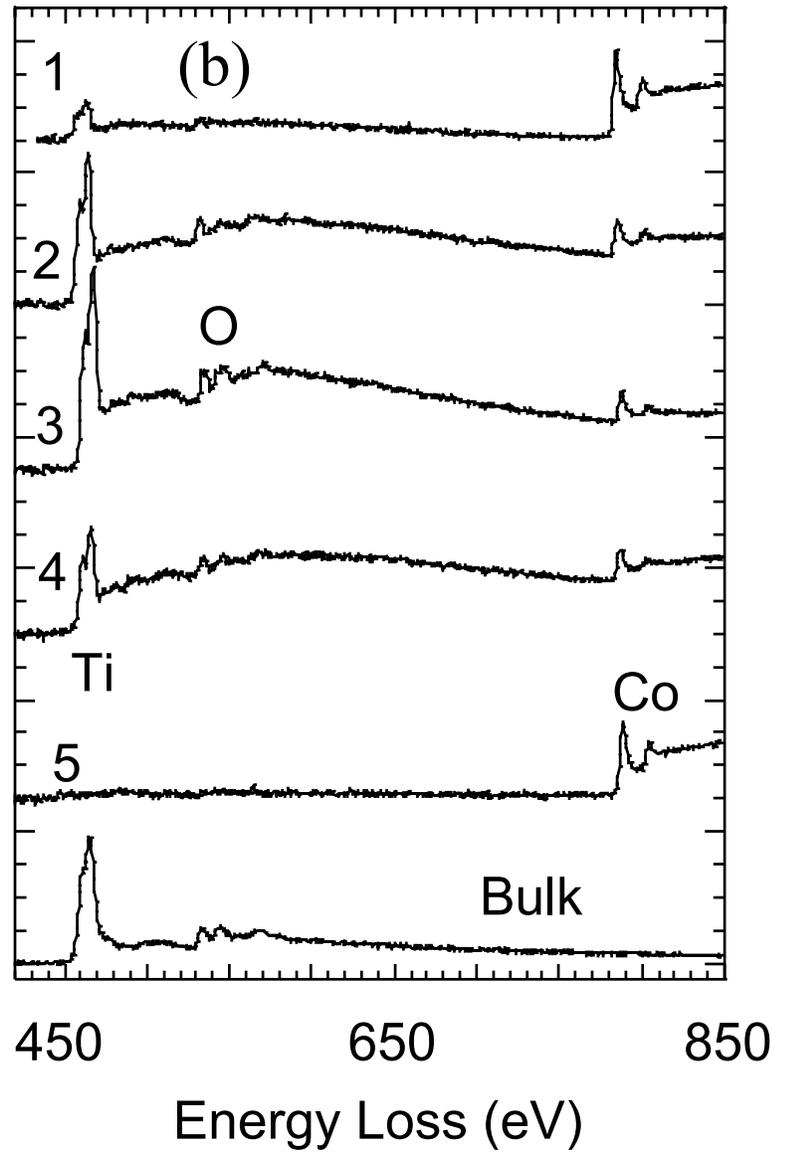

Fig. 4: S.R. Shinde *et al*.

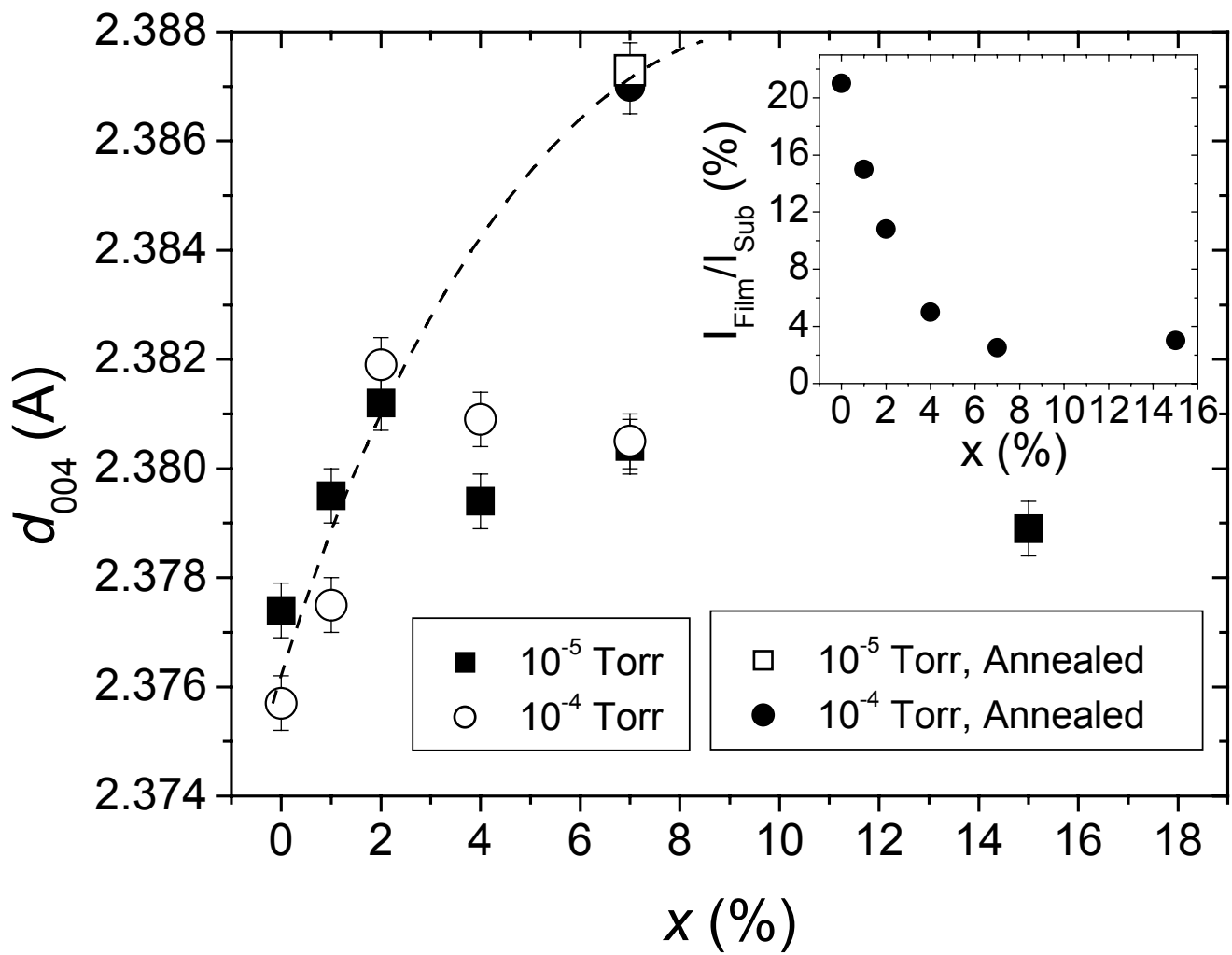

Fig. 5: S.R. Shinde *et al*.

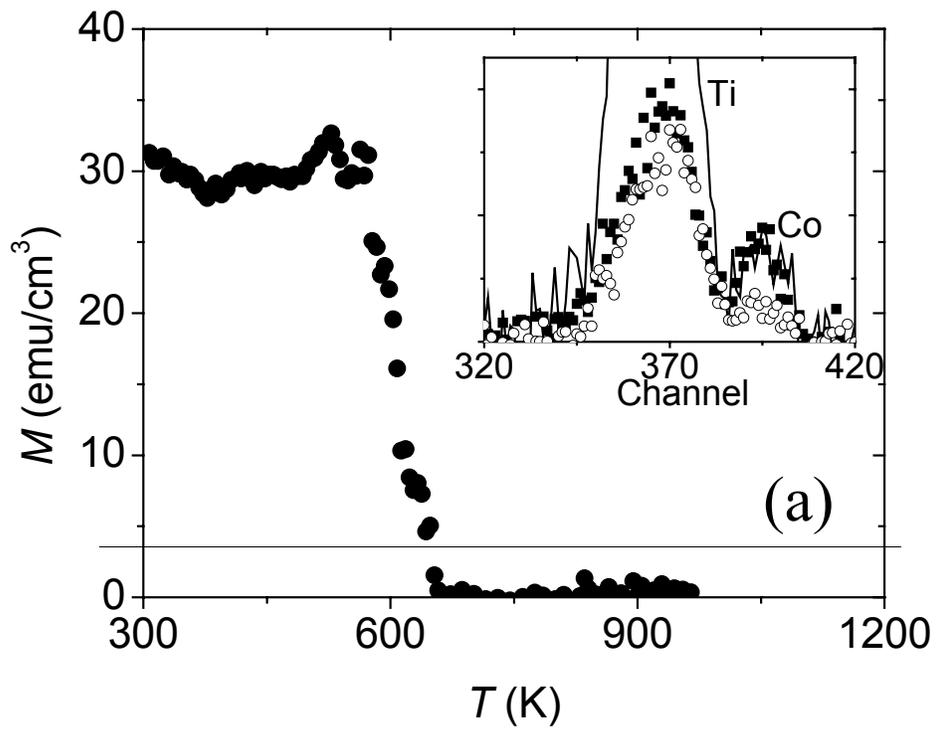

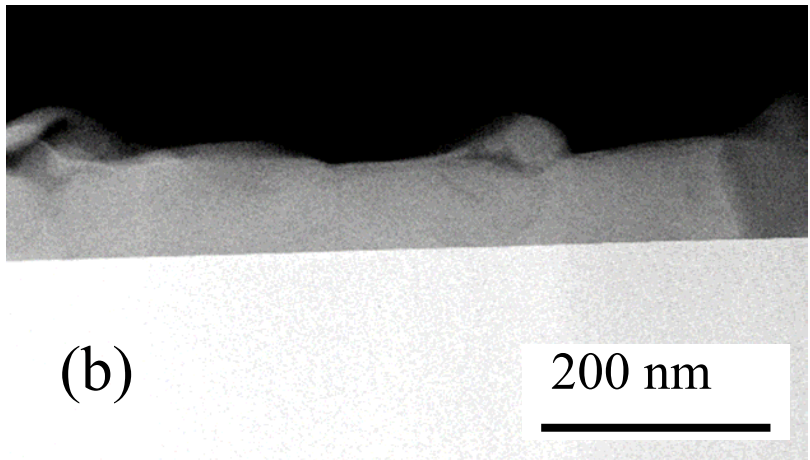

Fig. 6: S.R. Shinde *et al*.

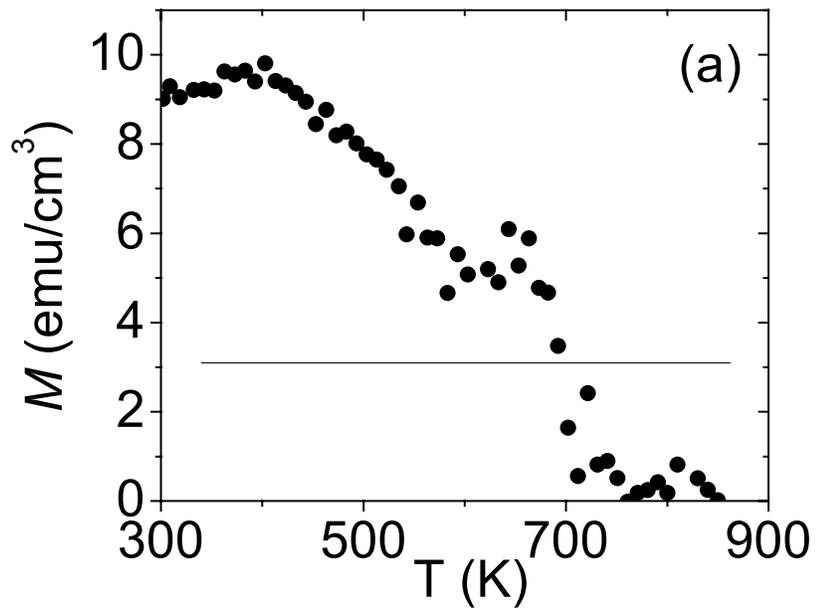

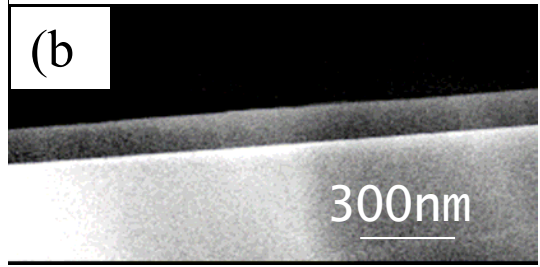

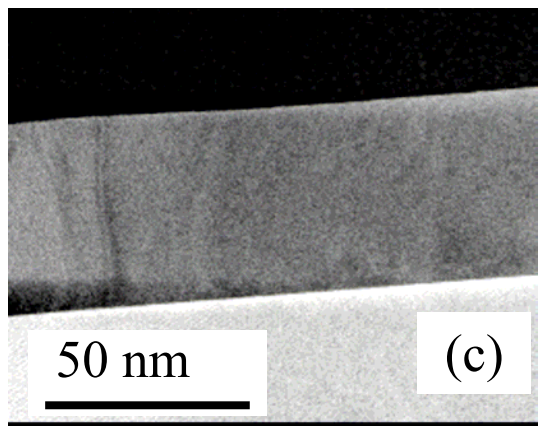

Fig. 7: S.R. Shinde *et al*.

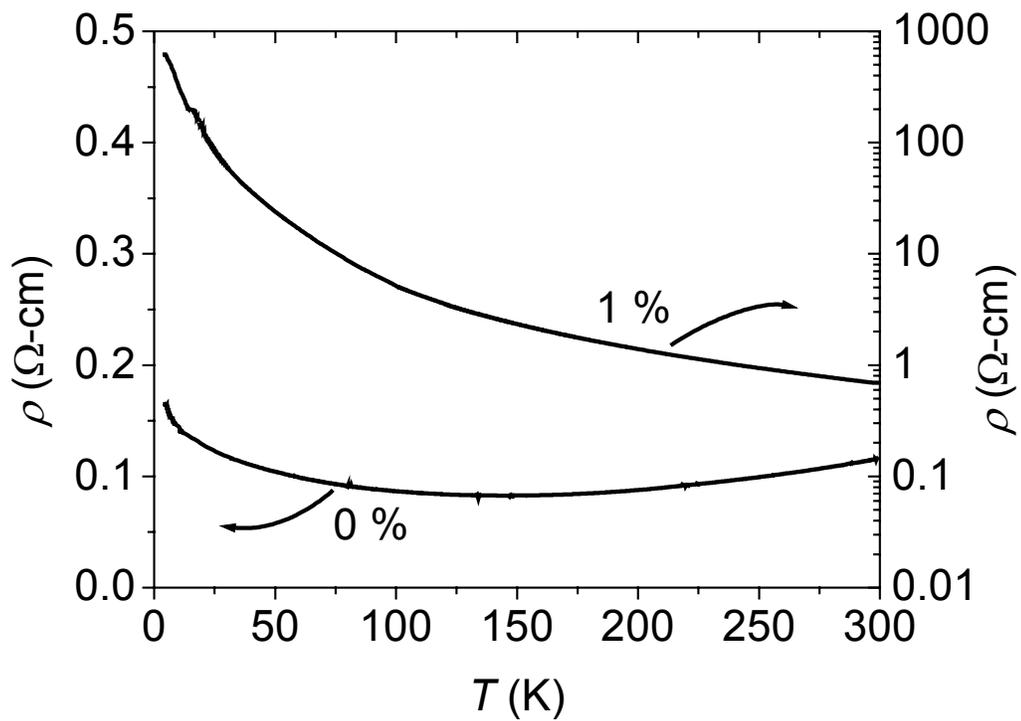
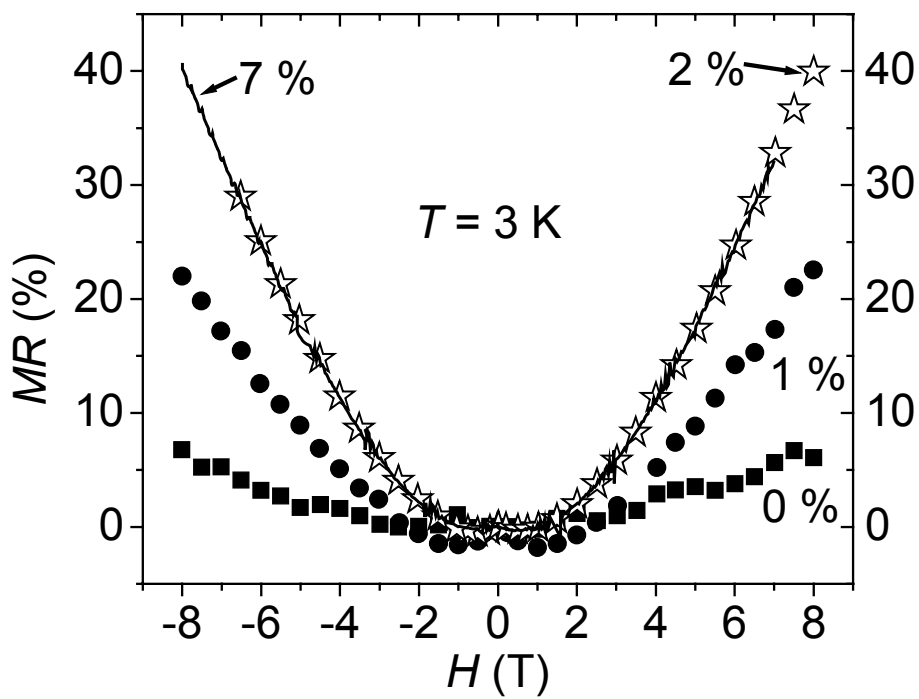

Fig. 8: S.R. Shinde *et al*.

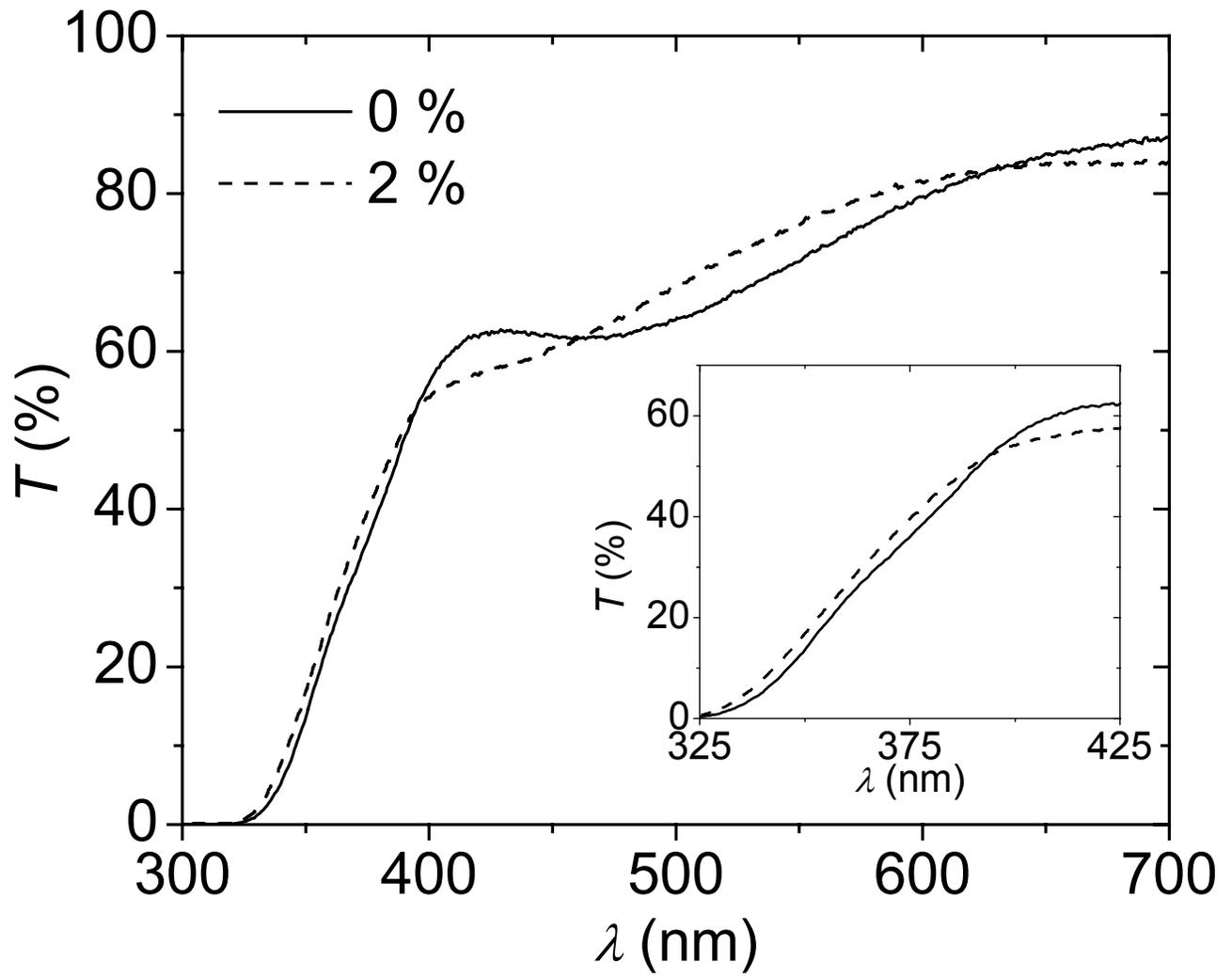

Fig. 9: S.R. Shinde *et al*.